Pre-publication version

*Assuring Autonomy International Programme*

# TIGARS

Towards Identifying and closing Gaps in Assurance of
autonomous Road vehicleS

Part 2

Tigars Technical Notes

Part 1

Assurance – overview and issues
Resilience and Safety Requirements
Open Systems Perspective
Formal Verification and Static Analysis of ML Systems

Part 2
Simulation and Dynamic Testing
Defence in Depth and Diversity
Security-Informed Safety Analysis
Standards and Guidelines

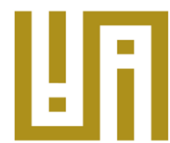

**ADELARD**

Pre-publication version

Editors & Author(s):
Robin Bloomfield, Adelard LLP and City, University of London
Gareth Fletcher, Adelard LLP
Heidy Khlaaf, Adelard LLP
Philippa Ryan, Adelard LLP

Contributing Author(s):
Shuji Kinoshita, Kanagawa University
Yoshiki Kinoshita, Kanagawa University
Makoto Takeyama, Kanagawa University
Yutaka Matsubara, Nagoya University
Peter Popov City, University of London
Kazuki Imai, Witz Corporation
Yoshinori Tsutake, Witz Corporation

# TIGARS

Towards Identifying and closing Gaps in Assurance of autonomous Road vehicleS

## TIGARS TOPIC NOTE 5: SIMULATION AND DYNAMIC TESTING
## Summary

Simulation has emerged as one of the most important means of assurance for Machine Learning (ML) embedded in control systems, but there are many challenges and areas of uncertainty surrounding its use. In this document we present a summary of issues as well as experience gained from the TIGARS project demonstrator in the autonomous vehicle domain.

24 Waterside
44–48 Wharf Road
London
N1 7UX

T +44 20 7832 5850
F +44 20 7832 5870
E office@adelard.com
W www.adelard.com

Authors
Philippa Ryan (Adelard)
Gareth Fletcher (Adelard)
Kazuki Imai (Witz Corporation)

### Use of Document

The document is made available as a resource for the community, providing all use is adequately acknowledge and cited.

This document provides a snapshot of work in progress. We welcome feedback and interest in this work. Please contact the authors or admin.tigars@adelard.com

### Acknowledgement

This project is partially supported by the Assuring Autonomy International Programme, a partnership between Lloyd's Register Foundation and the University of York. Adelard acknowledges the additional support of the UK Department for Transport.

## Contents

## Figures

## Tables

# 1 Introduction

Simulation has emerged as one of the most important means of assurance for Machine Learning (ML) embedded in control systems, but there are many challenges and areas of uncertainty surrounding its use. In this document we present a summary of issues as well as experience gained from the TIGARS project demonstrator in the autonomous vehicle domain.

# 2 Simulation and dynamic testing

Simulation is an approach widely used and encouraged, e.g., by the NHSTA [1], to train and verify the performance of ML used in autonomous vehicles. Simulation can be performed at many different levels of abstraction, some of which are described below:

- Fully virtual simulation – where the ML is executed in isolation with fully electronic input data and data capture. For example, running an image classification Convolutional Neural Network (CNN) on a PC with sample image file.
- Hardware in the Loop (HIL) – where the ML is run on representative hardware, however the inputs and outputs are managed virtually or in an artificial environment. For example, putting an autonomous vehicle inside a room with a bank of monitors and capturing decisions via data logging. Simpler cut down versions may also be used, e.g., a sub-system in isolation but with hardware sensors.
- Real-world limited trial – where the autonomous system is run on representative hardware but in a controlled environment, such as on a test track.
- Real-world trial – where the autonomous system is put into the public environment, with no control of test conditions.

Simulation may require substantial computer resources to create an environment with enough fidelity to gather meaningful results.

A serious assurance challenge is the amount of experience and testing needed in an autonomous vehicle to gain confidence. To match a human driver fatality rate of 2 - 3 per billion miles it is estimated that "*fully autonomous vehicles would have to be driven hundreds of millions of miles and sometimes hundreds of billions of miles to demonstrate their safety in terms of fatalities and injuries. Under even aggressive testing assumptions, existing fleets would take tens and sometimes hundreds of years to drive these miles — an impossible proposition if the aim is to demonstrate performance prior to releasing them for consumer use. Our findings demonstrate that developers of this technology and third-party testers cannot simply drive their way to safety.*" [2]. This is reinforced by Koopman in [3].

Therefore, simulation without resorting to real-world trials is seen as a practical way to gain assurance regarding the performance of an autonomous vehicle although it is an open question how his can be combined with other assurance evidence to give sufficient confidence in the safety of the system (the overall assurance is discussed in [4]). Additionally, only using real-world trials is not generally considered an ethical or a responsible choice, at least not without some reasonable assurance of safe performance before the vehicle is in contact with the general public and also for the occupants.

It should be noted that simulation discussions in this document are limited to simulation environments for verification and reinforcement learning of ML, rather than, for example, simulations of overall traffic flow once autonomy has been incorporated.

Table 1 below summarises the pros and cons of different combinations of virtual and real-world simulation. In practical terms it may be desirable to use different types at different stages of ML development. This would be dependent on the risk associated with the system, as that would inform the amount of evidence required to demonstrate adequate safety.

| Environment | ML | Strengths | Weaknesses |
|---|---|---|---|
| Virtual | Virtual | Can control and model many different environment options, which may be hard to replicate in real world testing<br><br>Can create accident sequences to test corner cases without risk of accident<br><br>Potentially cheap and quick<br><br>Can do early in lifecycle to assess performance<br><br>Can monitor every aspect of performance<br><br>Can use for reinforcement learning<br><br>Potentially strong repeatability<br><br>Easier to detect how/where faults occurred with monitoring | Unrealistic input data e.g., computer generated environment[1] or modelled sensor functionality which may not match the resolution and real-time performance of a real sensor<br><br>Extensive computer resources will be required to achieve the performance required for adequate modelling and collecting data e.g., in terms of processing power and fast access memory<br><br>ML may not perform this way in real life<br><br>Hard to involve user if needed<br><br>Potentially unrepresentative results (e.g., no feedback from bumpy surface, compromise of equipment from wet surface, temperature changes) |
| Virtual/Artificial | On target hardware (Hardware In Loop (HIL)) | Can control many different environment options<br><br>Can create accident sequences with very limited or no risk<br><br>Can involve end user<br><br>Gain trust in ML hardware<br><br>Potentially strong repeatability<br><br>Easier to detect how/where faults occurred with monitoring | Unrealistic input data – ML may be real but some of the input data may not be realistic e.g., if working in a room with lots of monitors<br><br>Computing power required may be large<br><br>Outputs may be more realistic but still constrained by environment (e.g., no actual movement or slower/faster responses)<br><br>User may not behave as they would in real environment or may have simulation sickness [5] |
| Real world but controlled e.g., test track | On target hardware | Input data is real and may contain unanticipated events<br><br>Can get useful feedback on performance with low risk to third party<br><br>Can involve users if needed | Less control over the environment<br><br>Much harder to repeat results<br><br>Harder to detect how/where faults occurred |

---

[1] Consider the situation where the simulation provides conflicting and unrealistic sensor data e.g., blocky low-resolution models, moving trees and unrealistically fast pedestrians [6]. Whilst it might be useful for the ML to identify this as invalid input data, if used for training care will be needed not to reinforce invalid behaviour.

| Environment | ML | Strengths | Weaknesses |
|---|---|---|---|
| Real world trials | On target hardware | Input data is real and may contain unanticipated events<br><br>Can involve users if needed | No control over environment<br><br>Riskier to third parties depending on mitigations in place<br><br>Hard to repeat results<br><br>Hard to detect how/where faults occurred |

Table 1: Simulation variants and their strengths and weaknesses

## 3   Simulation demonstrators

This section provides an overview of the simulation and dynamic testing performed on the TIGARS Evaluation Vehicle (TEV) golf cart, with an acceleration control system containing ML. The ML used was a version of the You Only Look Once (YOLO) [7] CNN which has been trained to detect people and vehicles, as well as other objects. The system under test uses a combination of distance calculations via parallax images, LiDAR and image classification to determine speed and acceleration settings. The system responds to other vehicles and pedestrians in its environment depending on their type(s) and distance from the vehicle.

### 3.1   Purpose

The purpose of our testing was as follows:

- verification of the effectiveness of existing testing methods for systems including ML models.
- elucidation of the gaps between actual and simulation environments for testing systems including ML models
- elucidation of the gaps between testing conventional systems (Non-AI) and systems including ML models

The tests were performed early in the development lifecycle of the system.

### 3.2   Environment

Tests were conducted in the following two simulation environments:

- TEV test room - Combination of Virtual and Artificial Environment with HIL
- Virtualized Verification into automatic Driving (ViViD) - Fully Virtual Environment

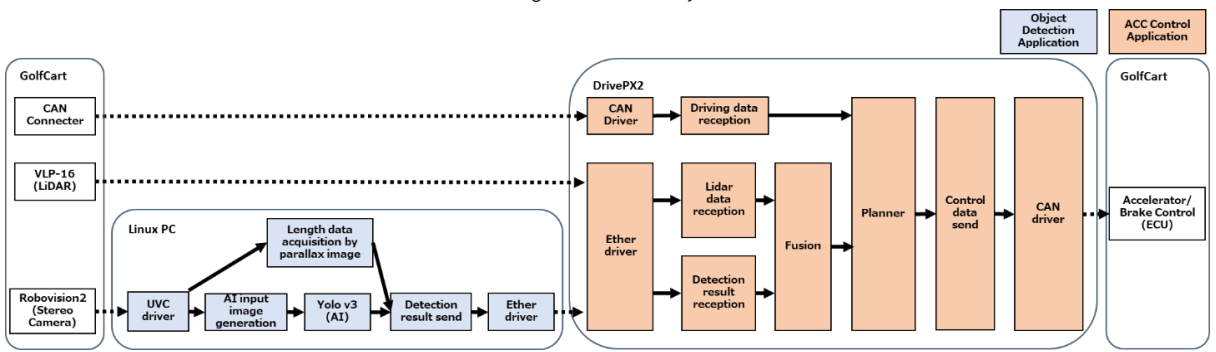

Figure 1: TEV test room configuration diagram

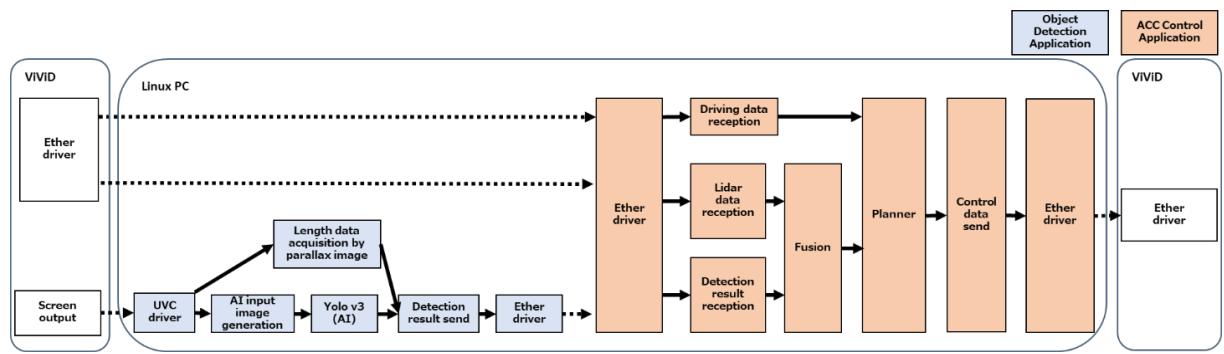

**Figure 2: ViViD configuration diagram**

### 3.3 TEV test room environment

These tests were conducted using a chassis dynamo. In the chassis dynamo environment, the TEV runs over the dynamo rollers. During the test, an environmental situation is reproduced by installing a panel of a person or a car in front of the golf cart. Since the space in which the chassis dynamo can be used is narrower than real life, the threshold values of the distance from the front vehicle when accelerating, decelerating, or stopping were adjusted proportionally.

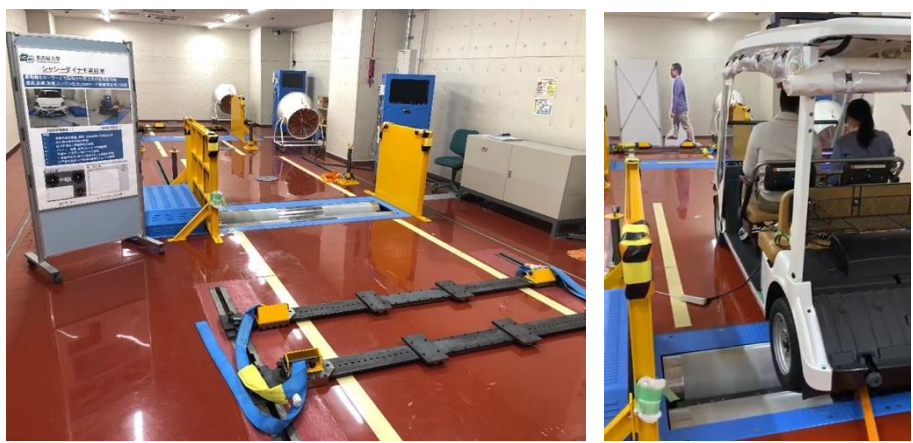

**Figure 3: Chassis dynamo environment**

### 3.4 ViViD environment

ViViD provides a fully virtual simulation environment for the TEV golf cart. Using ViViD, sensor information can be acquired by User Datagram Protocol (UDP) communication. It can be configured so that obstacles such as vehicles and pedestrians can be inserted into the environment, as well as failure injection within sensor data. The tests were carried out with the TEV driving on a typical road as shown below.

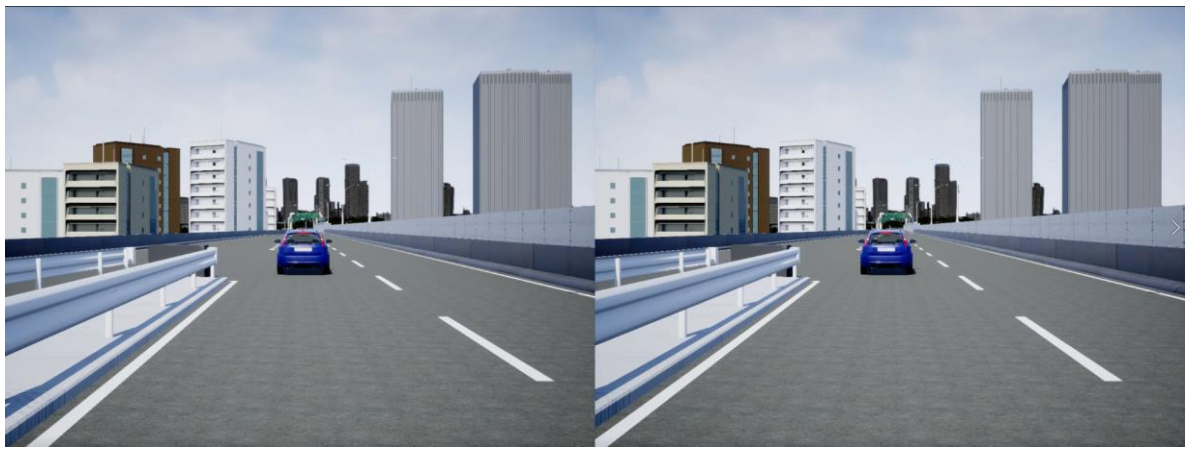

**Figure 4: ViViD environment**

## 4 Test cases and results

### 4.1 Test content

In order to verify whether the test methods were effective for the system including the ML model, test items were created as follows:

- normal scenarios were created based on system design specifications
- scenarios for failure, performance limitation, and misuse were created based on the results of safety analysis performed when the system concept was created

The created test items were further divided according to whether or not they could be tested in test room and ViViD environments, and the tests were carried out.

### 4.2 Lessons learned from running the experiments

This section describes the TEV experimental results and our analysis.

#### 4.2.1 Lessons from the ViViD environment

The system on the TEV uses comparisons of distance information from the object detection and LiDAR. There were many issues with timing in the ViViD environment which impacted on the effectiveness of the testing.

The LiDAR simulation software was too slow to be used at full fidelity in ViViD. As a result only part of the LiDAR data (the front ±15 degrees) was used to ensure a similar execution time as the real system. This was justified as it had no impact on the test cases being run within the ViViD environment.

A very highly specced machine was required to run the test application and simulator together, otherwise there were unacceptable delays sending the video output to the test application and in running the YOLO component. Even then, there were issues providing a predictable frame rate from the simulator, since only approximately 57 seconds of real-time data could be processed in around 1 minute. This had a cumulative effect on the simulation.

There were further complications with variations in the execution cycle, which changed from test to test. This meant that the tests were not repeatable. Only a rigid real-time execution of the simulator would have solved this, something that was impractical with off the shelf software. An attempt was made to lock-step time stamps from the LiDAR and object detection with the slowest input data, but the overall time lag meant that this was not a complete solution.

One important knock-on effect of the lack of repeatability is on regression testing. Tests cases re-run on a changed system cannot be assumed to execute with the changed functionality as the only variant, so the results of regression testing would need to be closely examined to ensure the results observed are valid and representative.

## 4.2.2   Results from the TEV test room environment

As the laboratory environment was small, the functionality of the software was adjusted in proportion to the size, i.e. the distance measurements and speeds were reduced. However, this had a knock-on effect on the overall response timing of the system which needs further analysis.

Some other issues arose as the TEV documentation did not describe all golf cart connectivity in detail, and a trial and error approach was needed to setup the CAN communication bus and to connect the required wiring.

To ensure that the autonomous control systems were working, the TEV was run with no object, within the braking distance (safety) range, for the vision systems to detect. The aim was to see if TEV would successfully accelerate to its target maximum speed and maintain that speed. The TEV did accelerate to its planned maximum speed and maintained this speed on the chassis dynamo; this showed us that the default planning behaviour of the TEV was working.

Then the autonomous system's responses to a vehicle being detected ahead at various distances were tested. Its behaviour should be adaptive, where the TEV target speed should reduce when the distance to the vehicle in front reduces too much allowing the vehicle in front to increase the gap between them before the TEV accelerates again to its target speed, and if the distance enters the safety region, the TEV should brake until it comes to a full stop. In practice we found that the detection rate of the vehicle was low, and the experiments with vehicle detection were almost impossible within the test room setup. During the experiments, a panel was used with the back of a car printed on to it to trigger the vehicle to stop, but the panel used had a lot of blank white space around the car (printed on the panel). This seemed to confuse YOLO and there were many cases in which the entire panel was recognized as a "bed" instead of a car. Occasionally, YOLO was sensitive to other extraneous items in the test environment as well (e.g., additional equipment was identified as "fridge"). The "bed" classification is most likely due to YOLO being trained on images with different context; for future experiments, we think removing the blank spaces or printing a car in context (on a road) for the background would make YOLO's detection rate more effective. However, "car" was often also detected in the images but with a lower confidence value, so an alternative is to remove the "bed" classification output from YOLO (leaving the neural network weights intact but taking the second or third result). This type of issue wasn't seen with the ViViD experiments and it shows the importance of performing a varied testing programme to find more unexpected results.

In terms of measuring performance, we found there were discrepancies in the distance measurements obtained when integrating the results from YOLO with the distance detection results from the RoboVision stereo camera, and this was more noticeable at a close distance. There was also a problem with consistency over adjacent frames, as even when the object was at a fixed, safe, distance, it was sometimes judged to be too close and a braking instruction was sent.

Currently the present golf cart behaviour means that it cannot be made to accelerate again until it stops completely after sending a single safety brake instruction, therefore it will always decelerate and stop, even if the target speed has increased after the brake command was sent. This highlighted there is an issue with the resilience specification, as a single event will cause the golf cart to spuriously brake. A proposed solution is for the system to be updated to only react after a number of brake signals are sent consecutively.

We found that small scale laboratory experiments were not easy and encountered problems we did not expect. For example, after scaling the parameters of the tests due to the small amount of laboratory space available, the TEV then experienced large variations in the acquired distance from the golf cart vision sensors. The sensors had had relatively high accuracy when detecting objects at a distance originally assumed in the TEV specification. However, as it is complex and expensive to prepare a testing environment that is very similar to the actual deployment environment (e.g., test tracks or large scale experiments) a 'good' simulator may be better suited in some cases and was still felt to provide value.

Another issue found was a case in which an obstacle in the blind spot range of the camera could not be detected by the LiDAR, or was detected with low distance accuracy. The LiDAR is part of a safety monitor for the TEV, identifying items in the camera's blind spot and overriding if safety distance is breached. To reduce detection issues in the test room, the installed obstacle was moved from its initial position so that the LiDAR could detect it and send the brake signal. However our analysis of LiDAR data log showed the TEV should have stopped with much earlier timing than it did in the tests, and so further to determine analysis potential causes of the issue is required.

It should be noted that the problems "detection rate of vehicle is low", "cannot accelerate again after sending a brake instruction", and "cannot detect obstacles in the blind spot range of the camera" did not occur in the ViViD environments and only became apparent in the lab testing. This highlighted the importance of performing small scale lab testing as part of the testing trials. These were not problems with the control algorithm (which was the focus of the ViViD testing) but instead were differences in the assumed environment and behaviour from the actual behaviour and supported environment of the COTS equipment.

### 4.2.3   Common findings

Understanding the correctness of the results was greatly improved by drawing the detection range on the input images. This was true of both the ViViD and chassis dynamo testing.

Both sets of tests highlighted problems with the parallax information being provided by the object detection software which had a large amount of dispersion, particularly with close objects and depending on how many objects were detected. This indicated improvements were needed to the distance calculation algorithms.

## 5   Conclusions and recommendations

Simulation has emerged as one of the most important means of assurance for ML embedded in control systems, but there are many challenges and areas of uncertainty surrounding its use. Different combinations of virtual and real-world simulations can be used and, in practical terms, it may be desirable to use different types at different stages of ML development. This would be dependent on the risk associated with the system, as that would inform the amount of evidence required to demonstrate adequate safety. The Tigars case study provides some insights into the pragmatic issues in using simulation on real projects in which an experimental vehicle was being built from off the shelf components integrated with bespoke software, by a sub-system developer.

The simulation studies on the project uncovered a lot of issues, many of which were unrelated to the ML but instead undermined confidence in the test environment and equipment. Hence, even if the test results are as expected it is not clear if we can trust them. Some of the findings are not new, for example, uncertainty in COTS equipment is a known issue, but this along with the combination of unproven ML technology with unproven testing methodology and equipment means establishing a compelling assurance case is additionally challenging.

The following recommendations are made from this work:

- Simulation can have many roles in the development and assurance lifecycle: the roles of the different simulation variants should be specified and justified.
- Confidence in the simulation environment needs to be established. In other words, how much we *can* trust it, and how much do we *need* to trust it. This will include confidence in any simulation software (in the quality of its construction), in the fidelity of the sensor data compared to real-life, and hence our trust in the results produced (both positive and negative). Although many tools are available off the shelf to support simulation, in our experience, they did not perform as anticipated (ViViD had many timing issues) and they may not have been developed to the quality traditionally expected for safety critical systems testing.
- Adjustments in system behaviour may be needed to accommodate the simulation environment and these will need to be justified so that test evidence can be used in the overall assurance cases.
- Additional findings should be sought from the test cases. The HIL testing uncovered an undocumented feature of the golf cart where it would come to a complete halt rather than allow a controlled slow down

# TIGARS

Towards Identifying and closing Gaps in Assurance of autonomous Road vehicleS

## TIGARS TOPIC NOTE 6: DEFENCE IN DEPTH AND DIVERSITY
## Summary

This TIGARS Topic Note discusses defence in depth and diversity for autonomous vehicles. We provide background on diversity and some guidance on the deployment of the use of defence in depth and diversity for these types of systems based on the case studies performed during the TIGARS project.

24 Waterside
44–48 Wharf Road
London
N1 7UX

T +44 20 7832 5850
F +44 20 7832 5870
E office@adelard.com
W www.adelard.com

Authors
Robin Bloomfield
Gareth Fletcher
Peter Popov

Copyright © 2020

### Use of Document

The document is made available as a resource for the community, providing all use is adequately acknowledge and cited.

This document provides a snapshot of work in progress. We welcome feedback and interest in this work. Please contact the authors or admin.tigars@adelard.com

### Acknowledgement

This project is partially supported by the Assuring Autonomy International Programme, a partnership between Lloyd's Register Foundation and the University of York. Adelard acknowledges the additional support of the UK Department for Transport.

## Contents

### Tables

### Figures

# 1    Introduction

This TIGARS Topic Paper discusses defence in depth and diversity for autonomous vehicles. We provide background on diversity and some guidance on the deployment of the use of defence in depth and diversity for these types of systems based on the case studies performed during the TIGARS project. The key message from a policy point and system/risk owner's point of view is that diversity is important and should be introduced systematically and explicitly in the system and development lifecycle. For the developer and system architect, there are many options to consider for the ML component including the use of real time ensembles, diverse training sets and different tool chains.

# 2    Defence in depth and diversity

Defence in depth and diversity are fundamental to achieving high levels of safety within complex systems. Diversity[1] is a key concept and diverse redundancy is needed to counter common cause failures and epistemic uncertainties. It is a sound and widely used design principle in safety critical applications. Lack of diversity was a key factor in the 2003 North American power blackout as non-diverse backup systems failed in the same way as the primary systems (p.60 [1]).

The key factor, which determines how beneficial "design diversity" is, is the *failure correlation* between "diverse" components. Ideally, when one opts for "design diversity" one hopes that simultaneous channel failures either do not occur at all or, if they do, they are rare. A number of studies, e.g. [4][9] with non-ML based software demonstrated that the gains from design diversity may be significant but are usually *significantly lower* than one may hope under the assumption that diverse components would fail (statistically) independently.

Some experimental results on the correlation of failures are shown in Figure 1 and Figure 2. Figure 1 is from a seminal Nasa funded experiment (data from Knight (1986)) that shows the improvement in the probability of failure of missile detection algorithm as the mean performance improves. The other (Figure 2) is from a software competition with many thousands of entrants and shows the reliability improvement of a diverse pair, relative to a single version (from Meulen (2008)). The horizontal axis shows the average probability of failure on demand of the pool from which both programs are selected. The vertical axis shows the reliability improvement from having a second algorithm.

The main message from these experiments is that on average one gets one or two orders of magnitude improvement in the probability of failure on demand by deploying diverse systems. One explanation for this is that independent designers and developers make similar mistakes because of the inherent difficulty of the problem that the algorithm is solving. The presence of these correlations and the non-independence of failures is a robust result, replicated across experiments sponsored by Nasa, the nuclear industry and others.

---

[1] Or diverse redundancy, "The presence of two or more systems or components to perform an identified function, where the systems or components have different attributes so as to reduce the possibility of common cause failure, including common mode failure". Diversity could result in different development lifecycles, different organisations, and different implementation technologies. The term "redundancy" denotes replicated, sometimes identical, systems or structures e.g. in protecting against fire by having identical systems located in different places.

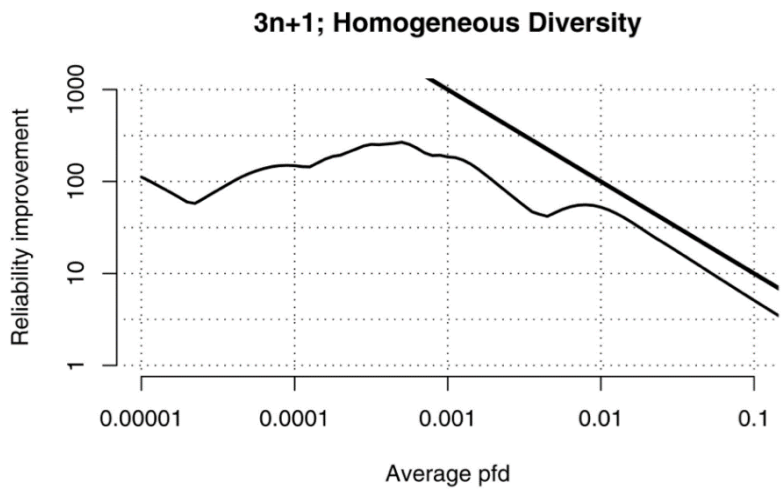

Figure 1: Experimental results on diversity

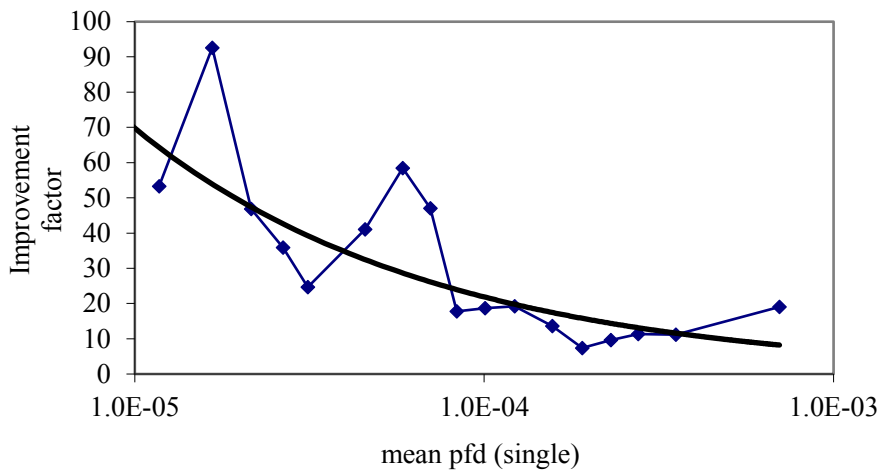

Figure 2: Experimental results on diversity improvement factor

Defence in depth in the autonomous vehicle context can take a variety of forms – from hardening a particular functional block (e.g. by deploying design diversity), to building a resilient architecture optimised to detect a failure, confine its impact and recover from failure fast. In addition diversity can be deployed within design and V&V teams, between development and assessment organisations, in tool chains to try and avoid problems of complex tool reliability and in V&V techniques [5].

The principles of how to deploy defence in depth are well-known and discussed widely in safety and security related standards and text books ([6][7][8]). For autonomous systems the challenge is how to deploy defence in depth with ML components. Such ML components may be used as "sensors" in a safety channel (e.g. to detect obstacles on the road) and also to implement an essential part of the functionality (e.g. in journey planners).

Diversity studies have been conducted with ML software, too. For instance, a number of studies in the late 1990s examined the effectiveness of design diversity with ML used for character recognition. In these works, e.g. [11], the authors made two observations:

1. The effectiveness of diversity is affected not only by whether diverse channels fail simultaneously, but also whether the failures are identical or not.

2. Diversity between channels can be promoted by carefully planning how the channels are trained, although the practical advice provided by the authors on how this can be done efficiently is very limited.

## 3 Defence in depth and diversity in TIGARS

The TIGARS project investigated the gaps and challenges for the assurance of autonomous road vehicles as a whole. Table 1 shows an extract from the gaps and challenges summary table [12] for defence in depth and diversity.

| Gaps and challenges area | Topic | Project response |
|---|---|---|
| Integration with defence in depth and diversity | Understanding how diversity and defence in depth can reduce the trust needed in specific ML components in the context of ML based systems. | Evaluate probabilistic models of resilience and defence in depth in the context of ML-based systems and the assurance case. Investigate the use of defence in depth and diversity in ML components and within the system architecture of RAS. |

**Table 1: Defence in depth and diversity: Gaps and Challenges**

TIGARS used two demonstrator systems for the defence in depth and diversity studies. The first is the TIGARS Experimental Vehicle (TEV), which is a modified Yamaha golf cart and has a use case of being a taxi on private property in which obstacle detection and adaptive cruise control are carried out by the installed autonomous systems. Figure 3 shows the physical golf cart after the installation of LIDAR (Light Imaging, Detection, And Ranging) and RoboVision camera test equipment. Secondly, Adelard and Nagoya have acquired Donkey car autonomous driving vehicles [13]. The Donkey car consists of the body of a Radio Control (RC) car, including motor and servo units, controlled by a Raspberry Pi computer and the Donkey car autonomous driving software (an open source python package using TensorFlow [14]).

### 3.1 Defence in depth and diversity studies on the TEV

The TEV has a typical Autonomous Vehicle (AV) architecture which we used to investigate some options for deploying defence in depth that are known to have been beneficial in other domains, e.g. sensors, processing information, algorithms etc. However, the assessment of the effectiveness of defence in depth is application specific and crucially depends on the correlation of failures between the diverse layers of defence.

The UML component diagram shown in Figure 3 captures a fragment of an architecture with ML components derived from the real architecture of the "golf car" (TEV), one of the case studies used in the TIGARS project.

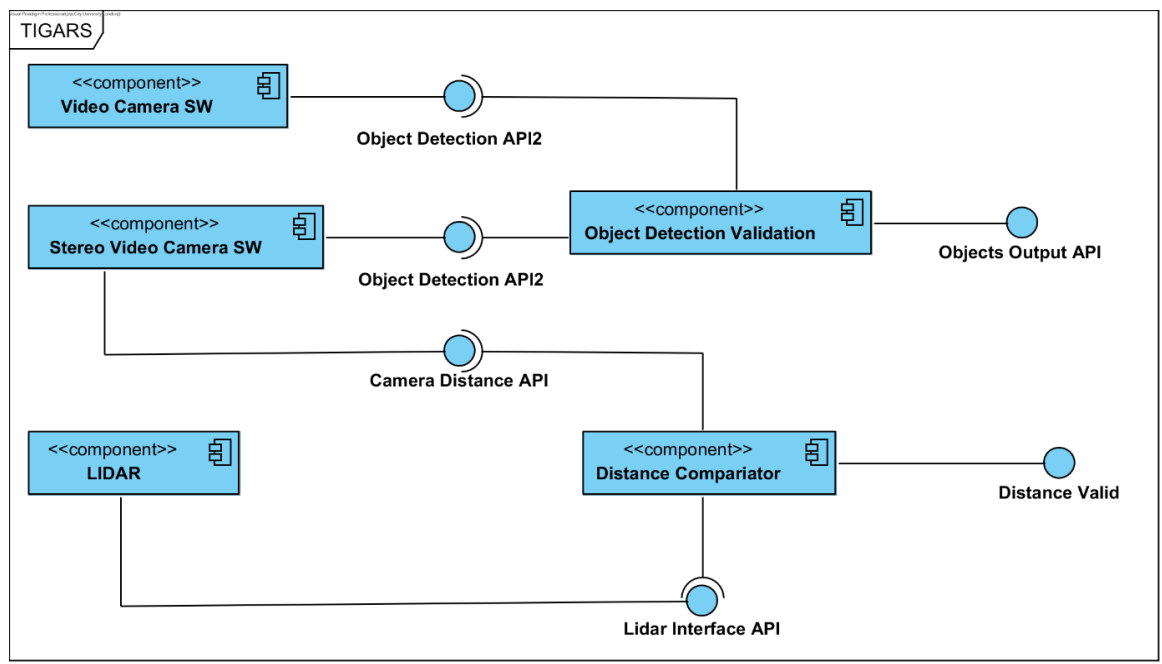

**Figure 3: Fragment of TEV architecture**

To improve reliability, both functions are implemented using "diverse" components (symmetric diversity); thus eliminating one type of common cause failure. Diversity in object recognition could be achieved by deploying two implementations of a CNN; two "functionally diverse" components are used for the distance measurement function too, one relying on the stereo camera as a sensor and the second on a LIDAR.

However, the two functions are clearly related (each of the channels implements the same functionality or the functionality of the channels is very similar), thus the outcomes from the two functions must be consistent: if objects are detected, the distance measurement should return a plausible value; if no objects are detected, the distance measurement function should return no value. In case of a disagreement between the channels the decision on which of the channels should be trusted is taken by an adjudicator, e.g. majority voting.

This is not possible in the TEV unless an additional channel is added or one of the two channels is trusted more than the other and the second channel is advisory (weakening the benefits of the diversity but still providing a checker/monitor). The TEV trusted the LIDAR distance information more as long as the object detection channels detected a vehicle and the stereo camera's distance information was used as a checker. Assessing the effectiveness of such an arrangement would need a detailed analysis of the failure correlation between the two channels: the effectiveness would only be undermined if there were circumstances in which the stereo camera would produce correct measurements while the LIDAR-based measurement would produce incorrect output. Less common examples of asymmetric systems, e.g. the LIDAR being used as a checker of an object recognition system based on a stereo camera, are not covered by [15], but the model can be refined to cover the specifics of the TIGARS architecture.

## 3.2 Neural network ensembles

Neural network ensembles (NNE) adopt "software design diversity" in neural networks. An NNE uses a finite number of individual neural networks for the same learning problem, and the final output is jointly decided by all the outputs of these individuals via an adjudicator.

Diversity is sought by:

1. diversifying the training data

2. diversifying the structure, the objective function used in training and/or even the type of the neural networks used in the ensemble

Broadly the ensembles are trained either in parallel ("bagging") or sequentially ("boosting"). A recent survey of the current state-of-the-art in NNE is given in [10].

As part of TIGARS, we also tested an asymmetric ensemble of models in our experimental trials with the Donkey car. A baseline ML model was used to perform an initial classification assessment - if the autonomous radio controlled car was on a straight or a corner part of the track. This is illustrated in Figure 4. Our initial results on the offline test bed showed a significant improvement over a single model approach. Although, we did notice some confusion factor with the classifier model, where it would send some cases to the wrong specialised model, (see [17] for more details on the studies).

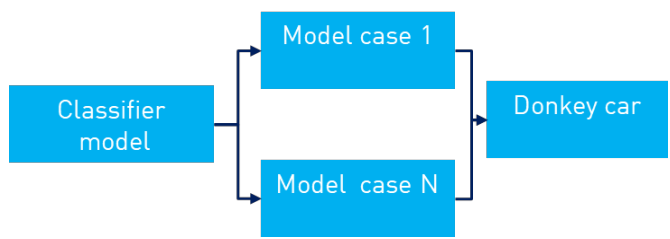

**Figure 4: Asymmetric diversity model architecture**

Then, we used one of the two more specialised models (one for the corners, another for the straights) to provide the output steering angle predictions. This type of asymmetric diversity is highly dependent on the classification model being able to differentiate between the two cases well, else there is confusion of which specialised model to use, and reduced accuracy and reliability if the incorrect model was used to make the output predictions.

In the AV context the work on asymmetric systems points to two important issues:

- The insight provided by [15] about the limitations of fault injection experiments may apply in the AV context.
- Some form of trusted checkers are an essential part of the "safety kernel" for them to be able to guarantee a high level of safety.

### 3.2.1 Regression faults in neural network ensembles

An interesting phenomenon, specific to Deep Neural Networks (DNNs), was observed empirically in the TIGARS project, which we called "regression faults"[2]. Regression faults occur with DNNs as a result of retraining (additional training) of a DNN; an Illustration is shown in Figure 5 below. The DNN is used to classify a communication "session" as either "clean" or "an attack".

---

[2] The name was chosen following the spirit of regression testing in software engineering, the focus of which is to eliminate "regression faults", which might be introduced in software in an attempt to fix other software faults. The manifestation of "regression faults" in software is part of software functionality that worked correctly before the attempted fix(es) for other software fault(s), but ceases to work correctly after the fix(es). Regression testing is meant to remove any "side effects" from the fault-fixing process.

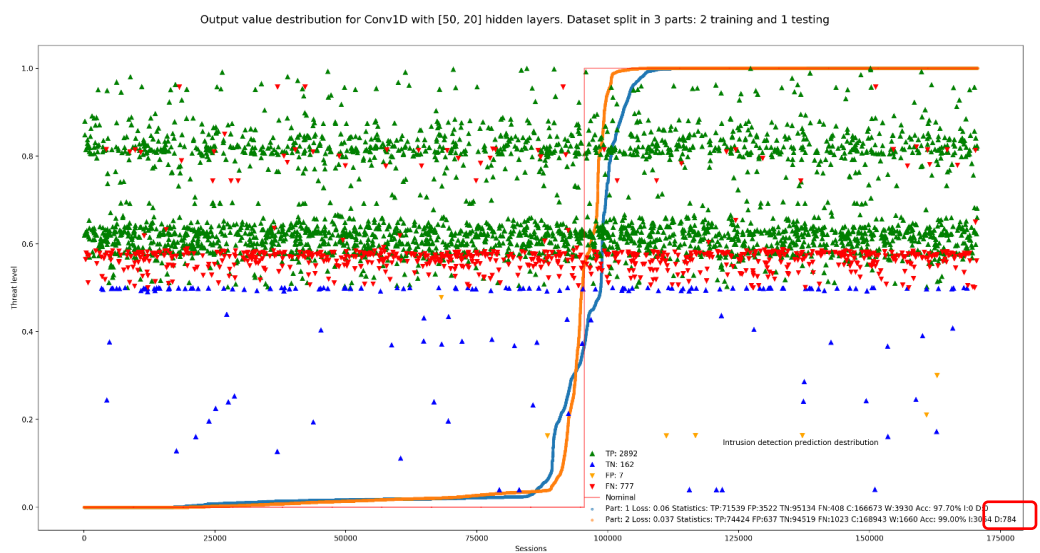

**Figure 5: An illustration of "regression faults" of a DNN subjected to retraining**

The plot represents the classification score of a DNN on the testing dataset of 170,603 sessions, which was derived from a much larger dataset of ~1,000,000 sessions, taken from the well-known CICIDS2017 dataset (https://www.unb.ca/cic/datasets/ids-2017.html). We split the initial 1,000,000 sessions into 3 parts: part 1 amounted to 40% of the data, and was used for initial training of the DNN; part2 also amounted to 40% of the overall data, and together with part1, was used to retrain the DNN after the initial training; part 3 used the remaining 20% of the data (i.e., the 170,603 sessions shown in Figure 5), and was used as a testing set to evaluate the DNN after the initial training (using part1 only) and also after DNN retraining (using both, part1 and part 2).

The classification scores on the testing data are ordered and plotted in Figure 5. The blue line shows the scores after the initial training; the orange line shows the scores after retraining. The accuracy, FP and FN rates are listed at the bottom of the plot. The accuracy of the DNN grows from 97.7% to 99% as a result of the retraining. "D:784", which is highlighted in the red box in the right bottom corner of the plot, indicates the manifestation of "regression faults": 784 items in the testing data set, which were classified **correctly** by the DNN trained on part 1 only, became classified **incorrectly** after the retraining (i.e., when the DNN was trained using both, part1 and part2). This is a very small proportion of the test cases <0.5%.

The existence of "regression faults" is not surprising - the nature of DNN is such that retraining does change the weights of the DNN, and in some cases may lead to changes of scores on different data items. The phenomenon has some specific implications for applying design diversity and defence in depth when some of diverse channels are ML-based. We point to two aspects, which seem important:

- "Regression faults" may affect the efficiency of the defence in depth. For instance consider that distance measurement in an autonomous vehicle is achieved by deploying two diverse channels - one based on an object recognition with a stereo camera and the other based on a LiDAR. Retraining the DNN to improve object recognition of the DNN may indeed lead to an improvement of object recognition. However, an undesirable consequence of the improvement may be the appearance of a "common failure" in the ML-based solution and the LiDAR if the DNN was covering for a particular class of failure in the LiDAR channel is no longer able to after the retraining. This possibility has not been studied in TIGARS, but the observations to date do not provide evidence to rule it out.

- If two or more channels of a diverse solution are ML based (i.e. if an ML-ensemble is used), then regression faults pose a new challenge for the verification of the ensemble. Our experiments indicate that retraining does lead to improvement of all properties of interest (accuracy, false negatives/positives) of a DNN ensemble on average, but we have established empirically that regression faults manifest themselves within ensembles too. Further details on these experiments with ensembles are provided in [17].

The important point here is that on average retraining improves scores, but "on average" might not be a sufficient measure in itself because for diversity evaluation we are also interested in changes to correlations. We note that theories are yet to emerge which either demonstrate that benefits from retraining are guaranteed to always outweigh the effect of "regression faults" or prove otherwise (i.e., that no such guarantees can be given in, at least, some cases or situations).

# 4    Recommendations

The key message from a policy point and system/risk owner point of view is that diversity is important and should be introduced systematically and explicitly in the system and development lifecycle. For the developer and system architect, there are many options to consider for the ML component including the use of real time ensembles, diverse training sets and different tool chains.

We make the following recommendations:

1. The use of diversity to improve reliability and safety is a sound principle. In particular it should be used to achieve higher dependability of safety mechanisms. The stakeholders for a mobility service or deployment of AVs should undertake a review of defence in depth and define a diversity and defence in depth strategy balancing the advantages of diversity with possible increases in complexity and attack surface.

2. Diversity should be considered in the system architecture to reduce the trust needed in a single ML component. Independence of failures should not be assumed and failure correlation should be considered based where possible on experimental data. For example, multiple sensors from different manufacturers should be deployed on independent channels within the autonomous vehicle.

3. There are a number of practicable ways in which diversity could be introduced into the ML lifecycle:

   - Software tools - different ML development platforms
   - ML model architectures and use of ensembles
   - Training data sets
   - Organisational diversity should also be considered with the ML development team independent from the testing and evaluation team.

4. The use of diversity to partition the operating regime (e.g. into areas with different types of difficulty) should be considered and the benefits of using ensembles and voting should also be evaluated.

5. Care should be taken when retraining DNNs to ensure that any regression faults do not pose new sources of potential failures for AVs post retraining. The average performance of the network may have improved; however, this could have been at the expense of introducing regression faults.

## Appendix A
## Recent work on difficulty and ensembles

NNE have been used extensively. "Demand difficulty" – the difficulty of executing a particular demand (data point submitted for classification, object recognition, etc.) – is a key concept in diversity as it underlines the observed correlation in failures. While in traditional software "difficulty" of a demand is rarely known, with the ML-solutions the difficulty can be readily available in a binary classification problem. For instance, Figure 5, clearly shows the scores of the individual demands, which vary between "1" (a clean communication session) and "0" (malicious communication session). The actual score from the classifier is an indication of demand difficulty: demands close to 0 will be confidently classified as "malicious sessions" and demands with scores close to "1" will be confidently classified as "clean sessions". We can see that a sizeable number of demands have scores which are not a clear cut - the ones which are "in between" with scores in the range [0.1, 0.9]. We explored the existence of difficulty and created an ensemble as follows: Observations about interpretation of the score from a binary classified as a measure of "difficulty" of demands have been made (see [16]), but we are not aware of a systematic use of "difficulty" (confidence). In TIGARS, we developed a boosting algorithm to create an ensemble consisting of several models, which may vary between two and an arbitrary number.

The ensemble is constructed as follows:

1.  Model 0 is trained on a training data set, T0, which includes 50% of the available data set. Post training, T0 is evaluated with the trained DNN to obtain the individual classification scores of the data points. We then split the T0 training set into two subsets for a predefined range of demand scores [a, b]: T0_easy - if a demand D has scores, $S(D) < a$ or $S(D) > b$ (within our confidence range). T0_difficult - If instead demand has score outside the range $S(D) \subset [a, b]$ (outside our confidence range). Intuitively, T0_easy contains the demands which are easier for the classifier to classify and T0_difficult contains the demands that are more difficult for the classifier to classify (i.e. the classification is less confident than those for the demands in T0_easy).[3]

2.  Model 1 is now trained using the T0_difficult data set. Intuitively, this model is trained to cope better with the cases which Model 0 finds difficult. This arrangement can be extended further, by splitting T0_difficult further into subsets which Model 1 finds easy or difficult to classify.

Eventually, we will have a set of models: Model 0, Model 1, Model 2, etc. which are meant to complement one another. The adjudicator of the models is based on the classification score, which is summarised in Table 2. The dashes indicate that that particular model does not see the input as it is not required to classify the demand, since it is within the confidence range of a previous model in the ensemble.

|  | Model 0 score | Model 1 score | Model 2 score | NNE output |
|---|---|---|---|---|
|  | $S_1(D) < a$ | - | - | 0 (Model 0) |
|  | $S_1(D) > b$ | - | - | 1 (Model 0) |
|  |  | $S_2(D) < a$ | - | 0 (Model 1) |

---

[3] The essential difference between the boosting algorithms, e.g. ADaBoost, Is that instead of using the scores of data on M0 to change the weights of the training data when training M0, we exclude the "easy" for M0 cases altogether from the training set used to train M1 and M1 will only be used when M0 fails to provide a confident classification.

| Demand, D | | $S_2(D) > b$ | - | 1 (Model 1) |
|---|---|---|---|---|
| | $S_1(D) \subset [a, b]$ (demand too difficult for Model 0) | $S_2(D) \subset [a, b]$ (demand too difficult for Model 1) | $S_3(D) < a$ | 0 (Model 3) |
| | | | $S_3(D) > b$ | 1 (Model 3) |
| | | | $S_3(D) \subset [a, b]$ | $S_3(D)$ (Model 3) |

**Table 2: Truth table for an adjudicator**

As one can see, the NNE output may come from any of the deployed models (M0, M1 or M2 in this particular example), depending on the "difficulty" of a demand for the respective individual models. If a demand is "easy" for Model 0, then Model 1 and Model 2 will not even see the particular demand. If Model 0 finds the demand difficult to classify, then Model 1 is called upon to classify the demand and based on the difficulty, Model 2 may even be called.

We found that the ensemble brought significant benefits over a single model using the CICIDS2017 to classify malicious network sessions.

# TIGARS

## TIGARS TOPIC NOTE 7: SECURITY-INFORMED SAFETY ANALYSIS

## Summary

This TIGARS Topic Note details the guidance for security-informed safety (SIS) analysis. We outline the issue for autonomous vehicles and provide some guidance on the deployment of the use of defence in depth and diversity for these types of systems based on the case studies performed during the TIGARS project.

24 Waterside
44–48 Wharf Road
London
N1 7UX

T +44 20 7832 5850
F +44 20 7832 5870
E office@adelard.com
W www.adelard.com

———————————————

Authors
Robin Bloomfield
Peter Bishop
Gareth Fletcher

Copyright © 2020

### Use of Document

The document is made available as a resource for the community, providing all use is adequately acknowledge and cited.

This document provides a snapshot of work in progress. We welcome feedback and interest in this work. Please contact the authors or admin.tigars@adelard.com

### Acknowledgement

This project is partially supported by the Assuring Autonomy International Programme, a partnership between Lloyd's Register Foundation and the University of York. Adelard acknowledges the additional support of the UK Department for Transport.

## Contents

### Tables

### Figures

# 1    Introduction

This document details the guidance for security-informed safety (SIS) analysis. We outline the issue for autonomous vehicles and provide some guidance on the deployment of the use of defence in depth and diversity for these types of systems based on the case studies performed during the TIGARS project.

# 2    Landscape

Security and safety have often been treated as separate disciplines, with their own regulation, standards, culture and engineering. Security requirements for vehicles are addressed in standards, such as PAS 1885 [1] and ISO 26262 [2], but not in an integrated way with safety, particularly the impact of functional safety requirements on security and the possible hazardous consequences from an attack or intrusion of the system.

This approach is no longer feasible as there is a growing understanding that security and safety are closely interconnected: it is no longer acceptable to assume that a safety system is immune from malware because it is built using bespoke hardware and software, or that it cannot be attacked because it is separated from the outside world by an "air gap".

In reality, the existence of the air gap is often a myth [see [4][5]]. Furthermore, autonomous systems rely on data and software with uncertain provenance and are not designed for high integrity applications. A safety justification, or safety case, is incomplete and unconvincing without a consideration of the impact of security.

The impact of cyber issues is exacerbated by the increasing sophistication of attackers, the commoditisation of low-end attacks, and the increasing vulnerabilities of digital systems as well as their connectivity – both designed and inadvertent. The cybersecurity compromises of automobiles have been steadily increasing, the Jeep attack and Tesla of a few years ago are no longer isolated incidents. Recently there has been an increased focus on attacking servers of the manufacturers that cars are becoming increasingly connected to. Such attacks are just examples of the type of adversaries a holistic safety case needs to address.

For example, the following areas are particularly significant from a security perspective and need more scrutiny in a security-informed justification of a safety system.

1.  Integration and interaction of requirements, e.g. of safety, with security and resilience supported by security-informed hazard analysis techniques.
2.  Supply-chain integrity, e.g. mitigating the risks of devices being supplied compromised or having egregious vulnerabilities.
3.  Malicious events post-deployment that will also change in nature and scope as the threat environment changes, and a corresponding need to consider prevention (e.g. implementing a risk-based patching policy) but also recovery and resilience.
4.  Weakening of security controls as the capabilities of the attacker and technology change. This may have a major impact on the proposed lifetime of installed equipment and design for refurbishment and change.
5.  Reduced lifetime of installed equipment as there is a weakening of security controls as attackers' capabilities and technologies change.
6.  Threats to the effectiveness and independence of safety barriers and defence in depth.
7.  Design changes to address user interactions, training, configuration, and software vulnerabilities and patching. These might lead to additional functional requirements for security controls.
8.  Possible exploitation of the device/service to attack itself or other systems and the need for confidentiality of design and deployment information.

| 9. | The trustworthiness and provenance of the evidence offered. |
| --- | --- |

**Table 1: Security-informed safety issues**

There are technical drivers to integrate security into safety analyses – because of the interactions and trade-offs that are necessary to consider. For example, at the requirements stage, we might need to consider the security aspects of the information flow policy when a plant is under attack, or if degraded plant conditions impact the safety. Another type of issue that we might need to consider at the architecture level is whether a highly critical third party component has sufficient security provenance given its supply chain. Safety assessment involves building trust with the supply chain, visiting their factories and assessing their culture: these are all aspects highly relevant to security as well as safety.

## 3    Security-informed safety in TIGARS

The TIGARS project investigated the gaps and challenges for the assurance of autonomous road vehicles as a whole. Table 4 in Appendix A shows the gaps and challenges for security-informed safety identified in TIGARS and our project response to them.

Our work on security-informed safety focused around the demonstrator systems. The TIGARS Experimental Vehicle (TEV), which is a modified Yamaha golf cart and has a use case of being a taxi on private property in which obstacle detection and adaptive cruise control are carried out by the installed autonomous systems, was used as a case study to apply PAS 11281 (see Section 3.1) and also an example for a security-informed Hazops (see Section 3.2).

## 3.1    Applying PAS 11281 to TEV

In 2018, Adelard developed a code of practice and a publicly available specification (PAS 11281 [3]) for security-informed safety in the railway and automotive sectors respectively. These documents capture and record our knowledge of best practice for security-informed safety in the form of concrete recommendations and guidance, with references to more detailed guidance and standards as appropriate.

PAS 11281: Connected automotive ecosystems – Impact of security on safety gives recommendations for managing security risks that might lead to a compromise of safety in a connected automotive ecosystem.

The PAS covers both the entire connected automotive ecosystem and its constituent systems throughout their lifetimes (including manufacturing, supply chain and maintenance activities). We focused on the application specific to autonomous vehicles as all levels of vehicle automation and autonomy are in the scope of the document. Security in the supply chain can be rather difficult for vehicles as they tend to be very complicated and involve many organisations; the PAS attempts to address this issue by promoting the "good cyber citizen" approach where everyone promotes good security practices in their products and the ecosystem as a whole becomes more secure.

The PAS clauses address

1. Security policy, organization and culture
2. Security-aware development process
3. Maintaining effective defences
4. Incident management
5. Secure and safe design
6. Contributing to a safe and secure world

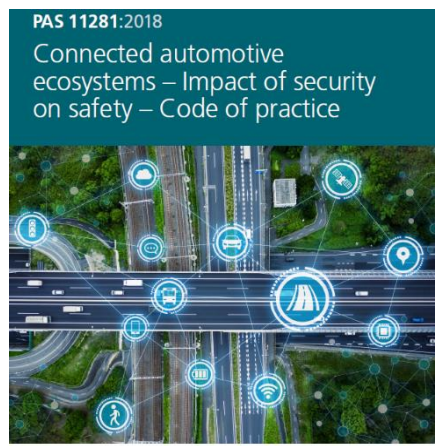

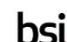

**Figure 1: Overview of PAS**

Although the findings of the case study showed that the vehicle development only addresses security in a preliminary manner, security is a fundamental and integral attribute to the technical themes of the project in the requirements, V&V, and assurance research. It was understood that security would be addressed more vigorously as the project matured during its life cycle.

## 3.2　Security-informed hazard analysis

One of the key topics in PAS 11281 is the impact of security on risk assessment covering the whole life cycle of the vehicle. The PAS states that security concerns could have an impact on:

1. the system boundaries;
2. what systems could potentially affect safety;
3. the stakeholders involved; and
4. the validity of design safety assumptions.

Therefore, care must be taken during the analysis to account for security concerns as well as safety. Table 2 summarises a 7-step risk assessment process: in TIGARS we applied Step 4 to the TEV.

| Step | Brief description |
|---|---|
| **Step 1 – Establish system context and scope of assessment** | Describe the system to be assessed and its relationship with other systems and the environment. Identify the services provided by the system and the system assets. Agree the scope of and motivation for the assessment and identify the stakeholders and their communication needs. Identify the type of decisions being supported by the assessment. |

| Step | Brief description |
|---|---|
| **Step 2 – Configure risk assessment** | Identify any existing analyses, e.g. safety cases, business continuity assessments that provide details of the system, the impact of failure and the mitigations that are in place. Characterise the maturity of the systems or project and the key uncertainties.<br><br>Ensure that the risk assessment is focused on the kinds of threats that are of concern. Define possible threat sources and identify potential threat scenarios. Refine generic capability and impact levels for the systems being assessed. Identify risk criteria.<br><br>Refine and focus system models in the light of the threat scenarios and existing analyses to ensure that they are at the right level of detail for an effective security-informed risk analysis. |
| **Step 3 – Analyse policy interactions** | Undertake an analysis of policy issues considering interactions between safety requirements and security policies. Resolve any conflicts, show that the trade-offs are satisfactory and document the decisions made. |
| **Step 4 – Preliminary risk analysis** | Undertake architecture based risk analysis, identifying potential hazards and consequences and relevant vulnerabilities and causes together with any intrinsic mitigations and controls. Consider doubts and uncertainties, data and evidence needs. Identify intrinsic and engineered defence in depth and resilience. |
| **Step 5 – Identify specific attack scenarios** | Refine preliminary risk analysis to identify specific attack scenarios. Focus on large consequence events and differences with respect to the existing system. |
| **Step 6 – Focused risk analysis** | Prioritise attack scenarios according to the capabilities required and the potential consequences of the attack. As with the previous step, the focus is on large consequence events and differences with respect to the existing system. |
| **Step 7 – Finalise risk assessment** | Finalise risk assessment by reviewing implications and options arising from focused risk analysis. Review defence in depth and undertake sensitivity and uncertainty analysis. Consider whether the design threat assumptions are appropriate. Identify additional mitigations and controls. |

**Table 2: 7-step security-informed safety risk assessment**

There are a variety of initiatives to integrate security into hazard analyses. We have been using security- (or cyber-) informed Hazard analysis and operability studies (Hazops) [8] to assess architectures of industrial systems [10], and adapt this well-known approach for performing a safety hazard analysis in a systematic fashion [9], analysing the deviations of data flows and values between different interconnections in the system. To account for security in a security-informed Hazops, additional security guidewords are added and an enhanced multidisciplinary team (system safety and security experts) is used. Both security and safety perspectives are needed to assess the likelihood of vulnerabilities being exploited and the effectiveness and consequences of their mitigations.

During TIGARS, we also performed a security-informed Hazops on the TEV architecture. This process is similar to the Hazops safety analysis with the addition of malicious security acts included in the possible causes of a hazard. We used a standard set of data flow and data value guidewords and reviewed key components of the architecture to understand the potential hazards in the system. The credibility and likelihood of a successful attack on the system depends on the capability level of the threat actor. We decided to consider threat actors with sophisticated capability and expert knowledge of the system. After all, once the vehicle is available for purchase there is nothing stopping a would be adversary from purchasing a target vehicle to acquire detailed knowledge and have a test bed for their attacks.

Figure 2 shows the simplified architecture that was used for the security-informed safety Hazops of the TEV. We focused on the interfaces which involved machine learning components, such as Object Detection and Fusion (denoted as 1, 2 and 3 in Figure 2). These components have additional complexity and differ from traditional components in road vehicles. It should be noted that the TEV is a research and development vehicle and not developed to any automotive standards.

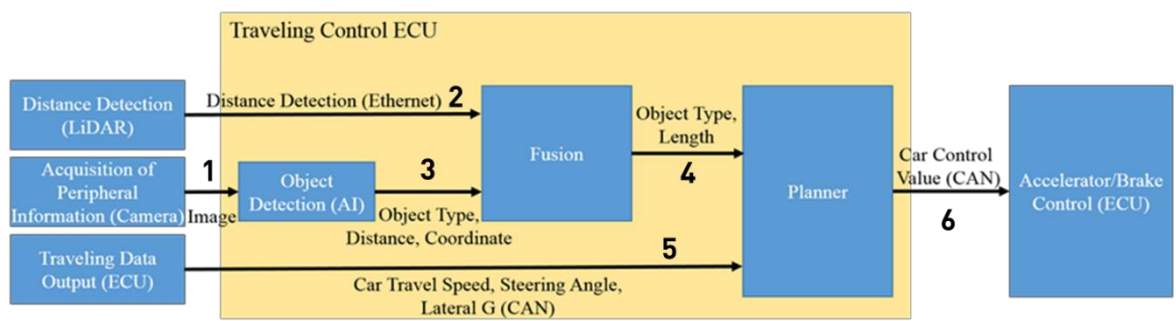

Figure 2: Overall architecture of the TEV

We found that security issues could pose credible threats to ML components if the inputs or outputs were able to be modified by the threat actor. We would expect real world autonomous vehicles (AVs to be more mature systems with additional security hardening than the TEV in our case study; however, security should still be considered during the risk assessment and design of the AV. Our hazard analysis highlighted some additional alarms and monitoring that could be added to the TEV to help annunciate potential failures and problems of the ML components.

An example extract from the hazard analysis summary for component 1 is shown in Table 3.

| Guideword | Interruption | Causes | Hazard | Mitigations |
|---|---|---|---|---|
| Data flow: No action | No image from camera | C1: Hardware failure<br><br>C2: Lens tampering | H3: Spurious safety stopping | M1: LIDAR cross-check<br><br>M2: Pre-test checks<br><br>R1: Diagnostic for camera feed failure<br><br>R2: Diagnostic check for image quality |

Table 3: Extract from hazard log summary of TEV for data flow 1

Table 3 shows a traditional hardware reliability cause with a more security focused cause both having possible contributing factors to a hazard. From this record in the Hazops, we made the recommendations that diagnostic checks should be added to check that the camera feed is alive and assess the quality of the image from the camera.

The hazard is because upon failure of the ACC the TEV will enter into an emergency stop procedure. Having this function activated too often represents a hazard for the system.

The other more traditional components without ML in the system are still susceptible to security compromise; for example if falsified/altered data was sent to the planner setting target speed it would be possible to crash the TEV into obstacles that the LIDAR sensors had detected, or even spuriously apply the emergency brake at opportune moments; the centre of a traffic junction could be a hazardous place to stop.

# 4 Guidance for security-informed safety in autonomous vehicles

Overall security-informed safety is not generally explicitly addressed in current AVs, and hence, the motivation for PAS 11281, let alone in a prototype vehicle being studied. However, security requirements may be partially met in mature implementation of the vehicle being studied. Overall, we consider the PAS will be challenging for industry. The results from applying the PAS in our case study may have been different if the TEV was not partly a research vehicle and a more mature system was being developed.

The deployment of autonomous technologies may follow an innovation cycle that first focuses on functionality and seeks to progressively add additional assurance and security. This will make the development of the assurance and safety cases and associated security and safety risk assessments particularly challenging. From our experience with the project we currently recommend:

1. Explicitly define the innovation cycle and assess the impact and feasibility of adding assurance and security. Adapt the 7-step risk assessment process to the specific lifecycle being used.

2. Address the approach to security-informed safety at all stages of the innovation cycle, including undertaking a security-informed hazard analysis during development. The hazard analysis should be reviewed periodically during operation or when a safety related component has been updated or additional threat or vulnerability information becomes available.

3. If safety, security and resilience requirements are largely undefined at the start of the innovation cycle, the feasibility of progressively identifying them during the innovation cycle should be assessed, together with the issues involved in evolving the architecture and increasing the assurance evidence.

4. Apply PAS 11281 to systematically identify the issues. If this is not possible because of the lack of defined processes or availability of information, consider a partial and project specific implementation of the PAS to meet the innovation cycle.

5. Collect experience in developing a security-informed safety case and in integrating security issues into the safety analyses needed to implement the PAS. In the industry as a whole, we believe that more training and expertise for SIS analysis is required, as many decisions rely on expert judgement, but the methodology that has been developed in other sectors is applicable to autonomous vehicles.

## Appendix A
## Gaps and challenges for SIS analysis

| Gap and challenge area | Topic | Project response – next steps |
|---|---|---|
| Security needs to be addressed throughout the lifecycle | Governance and policies on project rather than institutional basis.<br><br>The approach needs to address the innovation cycle. | Define an innovation cycle and research feasibility of progressive application of PAS. |
| Assurance cases and risks analysis | Need to integrate security into the safety case.<br><br>Need for analysis techniques that integrate security and safety. | The TIGARS CAE based assurance framework allows this to be done in principle. Develop and review framework after application.<br><br>Review and consider applying the security-informed safety case process currently being developed to one of the experimental vehicles.<br><br>This includes a risks analysis approach. Work with partners on understanding the differing approaches of STPA-SEC and security-informed Hazops. |
| Supply chain | Provenance of supply chain a generic issue. | In TIGARS consider the following ML issues:<br>Supply chain of ML based systems<br>Training data<br>Open source ML systems<br>Complex tool chain |
| Composition of cases | How to compose systems and assurance of heterogeneous COTS subsystems. | Specific to security aspects of these types of systems. |

Table 4: Security-informed safety: Gaps and challenges

# TIGARS

Towards Identifying and closing Gaps in Assurance of autonomous Road vehicleS

# TIGARS TOPIC NOTE 8: STANDARDS AND GUIDELINES
## Summary

This document gives an overview of International Standards and guideline documents relevant to assurance of RASs. This is a snapshot; the landscape is changing quickly as a number of activities have only started in the past year or two and are going to produce documents shortly. The documents are classified into three groups: on systems assurance including system life cycle processes, on AI and ML in general and on RASs. We also present our recommendations based on the review.

24 Waterside
44–48 Wharf Road
London
N1 7UX

T +44 20 7832 5850
F +44 20 7832 5870
E office@adelard.com
W www.adelard.com

Authors
Makoto Takeyama
Shuji Kinoshita
Yoshiki Kinoshita
(Kanagawa University)

Copyright © 2020

## Use of Document

The document is made available as a resource for the community providing all use is adequately acknowledge and cited.

This document provides a snapshot of work in progress. We welcome feedback and interest in this work. Please contact the authors or admin.tigars@adelard.com

## Acknowledgement

This project is partially supported by the Assuring Autonomy International Programme, a partnership between Lloyd's Register Foundation and the University of York. Adelard acknowledges the additional support of the UK Department for Transport.

## Contents

## Figures

## 1    Introduction

This document gives an overview of International Standards and guideline documents relevant to assurance of RASs. This is a snapshot; the landscape is changing quickly as a number of activities have only started in the past year or two and are going to produce documents shortly. The documents are classified into three groups: on systems assurance (Section 2) including system life cycle processes, on AI and ML in general (Section 3) and on RASs (Section 4). Our recommendations are presented in Section 5, and references are in Section 6.

## 2    International Standards on systems assurance and related documents

Top level International Standards on systems assurance and related documents are developed by ISO/IEC JTC 1/SC 7 *Software and systems engineering* [64] and IEC TC 56 *Dependability* [65]. The former develops and maintains the International Standards on assurance and system life cycle processes and the latter is responsible for standards on dependability.

The International Standard on systems assurance is ISO/IEC/IEEE 15026 *Systems and software assurance* that consists of four parts: *Part 1 – Concepts and vocabulary* [3], *Part 2 – Assurance case* [4] (revision work to start in 2020), *Part 3 – System integrity levels* [5] and *Part 4 – Assurance in the life cycle* [6] (revision work in progress). The Part 4 provides guidance and recommendations for assurance of a given claim about the system-of-interest. The guidance and recommendations of Part 4 are given for life cycle processes of the system-of-interest, rather than for the system, because of the need to support a claim of type "*The deployed system will continue to perform as required in future*", as recommended in the TIGARS topic paper on assurance [7]. Such a claim is itself about the system-of-interest, but typically derives descendant claims on the life cycle of the deployed system. Note that the definition of the term *assurance* in the AAIP BoK [66] and that in ISO/IEC/IEEE 15026-1 [3] differs: the former defines assurance as justified confidence while the latter defines it as grounds of confidence.

For the definition of the set of system life cycle processes, ISO/IEC/IEEE 15026-4 normatively refers to ISO/IEC/IEEE 15288 *System life cycle processes* [1] and ISO/IEC/IEEE 12207 *Software life cycle processes* [2], which are augmented by the multi-part guideline standard ISO/IEC/IEEE 24748 *Life cycle management*; in particular, its Part 1 [10] and Part 2 [11] contain the clarification of some concepts in [1] and [2]. The set of information items (documentations) relevant to each system life cycle process provided by [1] and [2] is identified by ISO/IEC/IEEE 15289 *Content of life cycle information items (documentation)* [9], which can be used as evidence in assurance arguments for the system-of-interest and in derived arguments for its life cycle.

It is often appropriate to consider an RAS as a System of Systems (SoS). Three International Standards ISO/IEC/IEEE 21839 *System of systems (SoS) considerations in life cycle stages of a system* [12], ISO/IEC/IEEE 21840 *Guidelines for the utilization of ISO/IEC/IEEE 15288 in the context of System of Systems (SoS)* [13] and ISO/IEC/IEEE *Taxonomy of systems of systems* [14] were published in succession in 2019; they together provide the basic concepts regarding life cycles of SoS referring to [1]. The four degrees of managerial and operational independence standardised by [14] à la Maier [15] may help clarify the complex structure of RASs.

IEC 60050-192 *International Electrotechnical Vocabulary (IEV) for dependability* [22] defines dependability as the "ability to perform as and when required". The term dependability is used as a collective term for the time-related quality characteristics of an item. It includes availability, recoverability, maintainability and supportability, and in some cases other characteristics such as safety, security and durability. As such dependability has a special role in assurance activities.

The top level standards of IEC TC 56 *Dependability* are IEV [22] and IEC 60300-1 *Dependability management - Part 1: Guidance for management and application* [21] (revision to start in 2020). The latter evolves the definition of dependability relevant concepts by providing guidance for management and application of dependability. IEC 62853 *Open systems dependability* [23] augments [21] with considerations for open systems (an open system is one whose boundaries, functions and structure change over time and is recognized and described differently from various points of view) of which RASs are naturally considered as instances.

The principal application of RAS assurance is safety. As a result of recent development of communication technology, safety and security are considered as inseparable with each other. BSI has recently published a new set of guidelines for security informed safety BSI PAS 11281:2018 *Connected automotive ecosystems. Impact of security on safety. Code of practice* [20].

Standards introduced in this section are depicted in Figure 1.

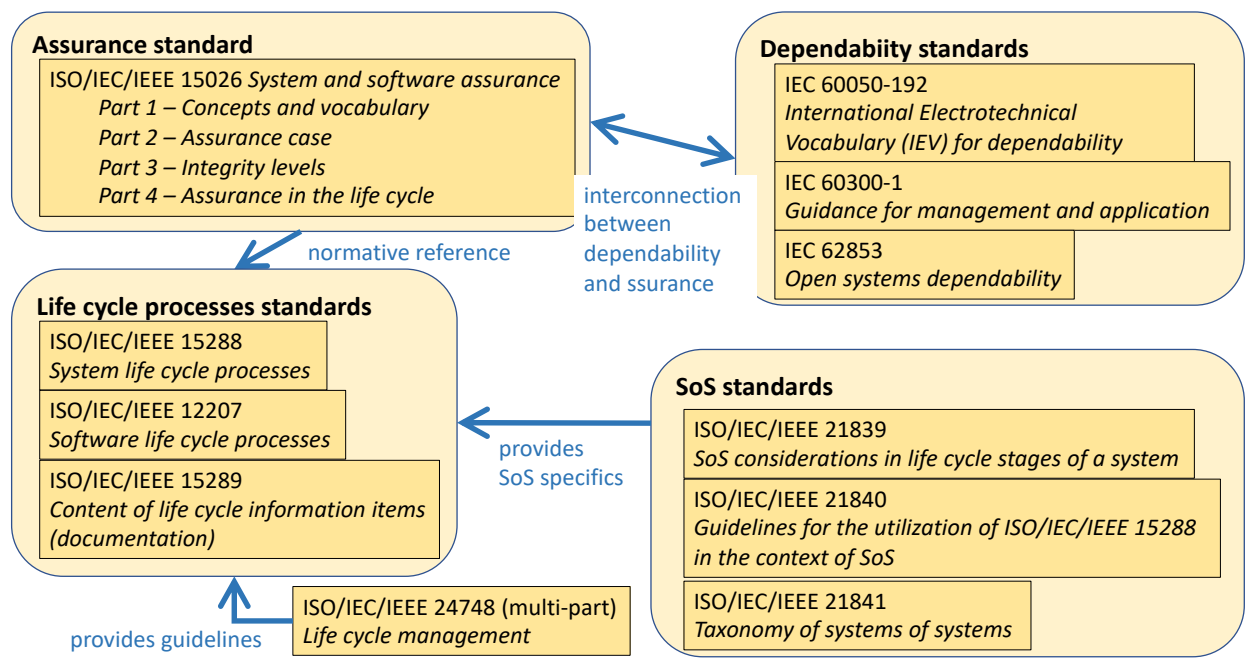

**Figure 1 International Standards related to assurance**

## 3   Standards and guidelines on AI

This section introduces standards and guidelines on AI in general, leaving RAS specific ones to Section 4.

OECD Council adopted a recommendation on AI [53] at Ministerial level on 22 May 2019 on the proposal of the Committee on Digital Economy Policy (CDEP). This adoption itself shows how high the impact of AI technology is recognised by OECD. Its recommendation consists of five Principles (Inclusive growth, sustainable development and well-being; Human-centred values and fairness; Transparency and explainability; Robustness, security and safety; and Accountability) and five National policies and international co-operation (Investing in AI research and development; Fostering a digital ecosystem for AI; Shaping an enabling policy environment for AI; Building human capacity and preparing for labour market transformation; and International co-operation for trustworthy AI). These lists are at a human-centred level including accountability, rather than at an engineering level.

The World Commission on the Ethics of Scientific Knowledge and Technology (COMEST) [54] is an advisory body and forum of reflection that was set up by UNESCO in 1998. It has published reports on extensive target fields that stems from space to water, including robotics [55]. It published Preliminary study on the Ethics of Artificial Intelligence [56]. IoT is included in its Work Programme 2018 – 2019.

ISO/IEC JTC 1/SC 42 *Artificial Intelligence* [59] was established in October 2017. As this group was only recently established, SC 42 has not yet published many International Standards except for standards on big data, but it has currently three Accepted Work Items (AWIs) for development of International Standards:

*Governance implications of the use of artificial intelligence by organizations* [24], *Process management framework for Big data analytics* [25] and *Risk Management* [26]. It also has the following AWIs and New Projects (NP) for development of technical report: *Overview of trustworthiness in Artificial Intelligence* [27], *Overview of computational approaches for AI systems* [28], *Overview of ethical and societal concerns* [29] and *Use cases* [30]. As for testing of AI systems in general, ISO/IEC JTC 1/SC 7/WG 26 *Testing* is developing ISO/IEC/TR 29119-11 *Testing of AI-based systems* [31].

There are study activities on ethics in AI and its applications by governments and International Standardisation bodies: the High-Level Expert Group on AI in European Commission, which presented *Ethics Guidelines for Trustworthy Artificial Intelligence* [39], ISO/IEC JTC 1/SC 42 WG 3 *Trustworthiness* [59], IEC SEG 10 *Ethics in Autonomous and Artificial Intelligence Applications* [32] and The IEEE global initiative for ethical considerations in AI and autonomous systems [60][61][62].

General guidelines for AI development and testing for improvement of AI systems quality can provide a basis for assurance arguments. Google's guidelines [35] describes recommended practices for developing AI systems based on their own experience of assurance in their projects. Google also provides a set of online ML development rules [36] based on their own experience. There are other RAS specific guidelines published by private sectors (See Section 4).

The StandICT.eu project report [53] provides a summary of international activities in AI from a more general viewpoint than this document.

## 4   Standards and guidelines on RAS

This section deals with standards and guidelines on RAS that employ AI technology. Standards and guidelines are introduced in two classes: requirements and testing (Section 4.1) and safety assurance (Section 4.2).

### 4.1   Requirements and Testing of RAS

There are some standards that instantiate general standards on requirements and testing to RASs and tailor them when necessary. For example, ISO 22737 [41], which is being developed by ISO/TC 204 Intelligent transport systems, is a standard for requirements and testing of Low-speed automated driving (LSAD) systems. Another example is IEEE P7009 [58] on Fail-Safe design requirements for RAS.

There are also attempts to introduce aspects that do not exist in the available standards. For instance, ISO/PAS 21448 [43] Safety of the intended functionality (SOTIF) for autonomous road vehicles developed by ISO/TC 22/SC 32 [63] plans to derive safety requirements from the functionality intended by the manufacturer, which is an aspect not covered by traditional safety standards, but is necessary for safety of RASs.

There are activities to establish forum standards on safety of autonomous driving systems. For example, Safety First for Automated Driving (SaFAD) [49] released by eleven mobility and automotive industry bodies, provides a framework for automated passenger vehicles. SaFAD is a non-binding organised framework for the development, testing and validation of safe automated passenger vehicles. Another example is the Association for Standardization of Automation and Measuring Systems (ASAM) who are developing a series of OpenX standards on file formats used for exchanging data in testing of autonomous vehicles, such as a format for logical description of road networks (OpenDrive [44]), a series of open file formats and open source tools for the detailed description, creation and evaluation of road surfaces (OpenCRG [45]) and one for the description of dynamic contents in driving simulation applications (OpenSCENARIO [46]).

Some research papers are relevant to building standards and guidelines on assurance of RASs. Requirements definition and validation of RAS is discussed by Koopman and Wagner [33], which suggests to prepare two sets of requirements: a set of ML training data, which accordingly pertains to ML elements, and another set of more traditional requirements; the two sets are to be used in parallel with a formal monitor that manages and restricts the ML output. Salay and Czarnecki [37] provide detailed analysis of the automotive ISO 26262 [19] standard regarding the software requirements and testing for machine learning systems; it proposes the use of tangible partial formal specifications, where possible, to provide plausibility checks, for example, pedestrian height or distance from the vehicle.

## 4.2 Safety assurance of RAS

For sector-wide certification, UL (Underwriters Laboratories) is developing *UL 4600: The First Comprehensive Safety Standard for Autonomous Products* [48]. The draft is under ballot. This standard has comprehensive guidance for achieving safety assurance of RASs by repeatable assessment of safety cases through a system life cycle.

There are national research projects in Germany and United States. The German research project PEGASUS aims for a model of scenarios with six independent layers and safety argumentation framework [50][51][52]. The DARPA's Assuring Autonomy project [34] aims for a methodology for building assurance cases using run-time monitoring of requirements, where an assurance case is built depending on and parameterized in conditional evidence which is given during development stage but is replaced based on the value measured with respect to the actual system and changing environment and updated according to the monitoring.

Some additional assurance frameworks are proposed. FiveAI published [40] intended to establish a basis for certification of Highly Automated Vehicles. They put emphasis on certification, verification and validation. There are attempts to form frameworks for RAS safety cases. The Safety Critical Systems Club (SCSC) is developing guidance [47] for safety of autonomous systems including a three-level framework: computational level (e.g. Route planning), autonomy architecture level (e.g. Sensor health checks, Sanity checks on generated route) and platform level (e.g. Self-driving car).

Uber Advanced Technologies Group's approach to the safe development of self-driving vehicle technology [38] is an internal guideline of Uber. It is a result of their self-review of safety approach reflecting their own experience of an accident.

## 5  Recommendations

1. Duplication of standardisation work on the same topic should be reduced to the minimum because it can result in inconsistency, as already observed in international standardisation activities for many years.

   We observe that some RAS relevant topics have multiple standards. An example of possible duplication is in risk management. ISO SC 262 [42] is devoted to risk management, and there is a planned standard in the artificial intelligence context [26] developed by ISO/IEC JTC 1/SC 42 and a published standard in the software and systems engineering context [16] developed by ISO/IEC JTC 1/SC 7.

   Another example is trustworthiness and dependability. ISO/IEC JTC 1/SC 42 has WG 3 trustworthiness and IEC TC 56 is devoted to dependability; the two concepts seem to be in close relationship.

2. An authoritative and introductory guideline covering necessary knowledge for the whole area of RAS should be developed for new entrants to this arena. Particularly, the guideline should include survey on foundational standards of the safety field.

   Many IT companies are entering into the market without the experience of the traditional manufacturers. The current lack of overall guidelines runs the danger that they will concentrate too much on their strength in a particular area without necessary knowledge. Because of the speed and scale of advancement of RAS engineering, many guidance documents are circulated in varying maturity at present. The recommended guideline would help ensure that their new technologies and traditional engineering fit together.

3. There are AI and ML specific issues that are particularly difficult to solve, such as testing of ML based system, treating human factors in the context of AI, adaptation of ML, treatment of learning data, and their mixture. Standards to help solving these issues should be developed, with priority over those that only make obvious specialisation where existing general standards would not solve the difficult issues.